# Persistent exchange splitting in a chiral helimagnet Cr$_{1/3}$NbS$_2$

Na Qin[1*], Cheng Chen[2*], Shiqiao Du[1*], Xian Du[1], Xin Zhang[3], Zhongxu Yin[1], Jingsong Zhou[1], Runzhe Xu[1], Xu Gu[1], Qinqin Zhang[1], Wenxuan Zhao[1], Yidian Li[1], Sung-Kwan Mo[2], Zhongkai Liu[3,4], Shilei Zhang[3,4], Yanfeng Guo[3,4], P. Z. Tang[5,6], Yulin Chen[1,3,4,7‡], and Lexian Yang[1,8‡]

[1]*State Key Laboratory of Low Dimensional Quantum Physics, Department of Physics, Tsinghua University, Beijing 100084, China.*
[2]*Advanced Light Source, Lawrence Berkeley National Laboratory, Berkeley, CA 94720, USA*
[3]*School of Physical Science and Technology, ShanghaiTech University and CAS-Shanghai Science Research Center, Shanghai 201210, China.*
[4]*ShanghaiTech Laboratory for Topological Physics, Shanghai 200031, China.*
[5]*School of Materials Science and Engineering, Beihang University, Beijing 100191, China.*
[6]*Max Planck Institute for the Structure and Dynamics of Matter, Center for Free Electron Laser Science, 22761 Hamburg, Germany.*
[7]*Department of Physics, Clarendon Laboratory, University of Oxford, Parks Road, Oxford OX1 3PU, UK.*
[8]*Frontier Science Center for Quantum Information, Beijing 100084, China.*

*These authors contribute equally to this work.
‡Email address: LXY: lxyang@tsinghua.edu.cn, YLC: yulin.chen@physics.ox.ac.uk

**Using high-resolution angle-resolved photoemission spectroscopy (ARPES) and *ab-initio* calculation, we systematically investigate the electronic structure of the chiral helimagnet Cr$_{1/3}$NbS$_2$ and its temperature evolution. The comparison with NbS$_2$ suggests that the electronic structure of Cr$_{1/3}$NbS$_2$ is strongly modified by the intercalation of Cr atoms. Our *ab-initio* calculation, consistent with experimental result, suggests strong hybridization between Nb- and Cr-derived states near the Fermi level. In the chiral helimagnetic state (below the Curie temperature $T_c$), we observe exchange splitting of the energy bands crossing $E_F$, which follows the temperature evolution of the magnetic moment, suggesting an important role of the conduction electrons in the long-range magnetic ordering. Interestingly, the exchange splitting persists far above $T_c$ with negligible temperature dependence, in drastic contrast to the itinerant ferromagnetism described by the Stoner model, indicating the existence of short-range magnetic order. Our results provide important insights into the microscopic mechanism of the chiral helimagnetic ordering in Cr$_{1/3}$NbS$_2$.**

## I. INTRODUCTION

Chiral materials lacking both inversion and mirror symmetries exhibit many interesting properties, such as chiral Weyl fermions and extremely long Fermi arc [1-6], quantized circular photogalvanic current [7], and chiral magnetic effect [8]. Particularly, if a chiral material hosts magnetic ordering, the competition between the ferromagnetic (FM) exchange interaction and Dzyaloshinskiie-Moriya interaction (DMI) can result in helical alignment of spins, forming the so-called chiral helimagnet (CHM). CHMs can harbour novel topological excitation with magnetic vortices known as skyrmions that can be effectively tuned by external magnetic field or spin-polarized electrical currents [9-12], promising great application potential for the spintronic devices. So far, CHM materials have been discovered in mainly two space groups: cubic space group $P2_13$, such as MnSi [9,13], $Fe_{1-x}Co_xSi$ [14,15], FeGe [16-18], $Cu_2OSeO_3$ [19,20]; and space group $P4_132$ or $P4_332$, such as β-Mn-type Co-Zn-Mn [21] and $FePtMo_3N$ [22].

Among the CHMs, $Cr_{1/3}NbS_2$ is unique since it hosts one-dimensional (1D) chiral soliton lattice instead of 2D or 3D skyrmions [23]. It crystallizes in a layered hexagonal structure with the space group of $P6_322$. Within each unit cell, there are two trigonal prismatic $NbS_2$ layers that are rotated by 180° with respect to each other. The Cr atoms occupy the octahedral holes between two adjacent $NbS_2$ layers and order in a ($\sqrt{3} \times \sqrt{3}$)R(30°) superstructure. $Cr_{1/3}NbS_2$ can therefore be considered as Cr-intercalated 2H-$NbS_2$ [Fig. 1(a)]. Below about 125 K, the intercalated Cr atoms with a local spin moment of about 3 $\mu_B$ order ferromagnetically and the system enters into a helimagnetic ground state with a large helix period of 48 nm along the chiral *c* axis [Fig. 1(b) and Fig. 1(c)]. When an external magnetic field is applied perpendicular to the *c* axis, a highly tunable and robust chiral soliton lattice can be observed [23], indicating the important application potential in spintronic devices.

Despite extensive research effort, the microscopic mechanism of the CHM ordering in $Cr_{1/3}NbS_2$ is still controversial. From the conventional understanding of magnetically intercalated transition metal dichalcogenides, the FM exchange interaction between Cr ions has a Ruderman-Kittle-Kasuy-Yosida (RKKY) form that is mediated by the itinerant Nb conduction electrons [24,25]. In this scenario, the local orbitals of Cr ions contribute negligibly to the Fermi surface (FS) of $Cr_{1/3}NbS_2$ but only dope electrons to the system. However, this picture has been challenged by recent experimental [26] and theoretical [24,26,27] results, which suggest significant contribution of Cr ions to the density of states near $E_F$. Indeed, previous angle-resolved photoemission spectroscopy (ARPES) experiments show strong hybridization between Nb- and Cr-derived electronic states near $E_F$ [28,29], and a Hund's exchange interaction, instead of RKKY interaction between Cr ions has been proposed [29]. To understand the mechanism of the novel magnetic ordering, it is essential to adequately study the electronic structure and its interplay with the magnetism of $Cr_{1/3}NbS_2$.

In this work, we systematically study the electronic structure of $Cr_{1/3}NbS_2$ and its temperature evolution using high-resolution ARPES. We reveal drastic difference between the electronic structures of $NbS_2$ and $Cr_{1/3}NbS_2$, consistent with our *ab-initio* calculation that suggests strong hybridization between Cr- and Nb-derived states near $E_F$. In the CHM state, we observe exchange splitting of energy bands crossing $E_F$, whose magnitude nicely follows the temperature evolution of the magnetic moment below the Curie temperature $T_c$, indicating strong coupling between itinerant conduction electrons and localized spin moments. Interestingly, the exchange splitting persists even at temperatures far above $T_c$, in drastic contrast to the itinerant FM described by the Stoner model. Instead, the overall temperature evolution of the exchange splitting can be well understood by the Oguchi model based on interacting localized spin moments.

## II. METHODS

The $Cr_{1/3}NbS_2$ crystals were grown via employing the chemical vapor transport method by using iodine as the transfer agent. Stoichiometric Cr powder (99.9%, Adamas), Nb powder (99.95%, Aladdin) and S powder (99.9%, Adamas) were mixed and grounded in a mortar. Then the mixture was sealed in a quartz tube together with the iodine under a vacuum of $10^{-4}$ Pa. The assembly was placed into a single temperature zone tube furnace and the temperature of the high temperature region was kept at 1000 °C. After two weeks, large crystals of $Cr_{1/3}NbS_2$ were obtained at the cold end of the quartz tube. The crystallographic phase and crystal quality examinations of $Cr_{1/3}NbS_2$ were performed on a single-crystal X-ray diffractometer equipped with a Mo Kα radioactive source ($\lambda = 0.71073$ Å). The diffraction patterns could be satisfyingly indexed on the basis of a $Nb_3CoS_6$ polytype structure (space group: $P6_322$, No. 182) with the lattice parameters $a = b = 5.74$ Å, $c = 12.09$ Å, α = 90°, β = 90° and γ = 120°.

High-resolution ARPES measurements were performed at beamline 13U of National Synchrotron Radiation Laboratory (NSRL), beamline 4 and beamline 10 of Advanced Light Source (ALS), and beamline 4.5.1 of Stanford Synchrotron Radiation Light Source (SSRL).

Data were collected with Scienta R4000 (DA30) electron analyzer at NSRL and ALS (SSRL). The overall energy and angular resolutions were set to 15 meV and 0.2°, respectively. The samples were cleaved *in situ* and measured under ultra-high vacuum less than $1.0 \times 10^{-10}$ mbar.

First-principles band structure calculations were performed using QUANTUM ESPRESSO code package [30] for the non-magnetic calculations and Vienna *ab initio* simulation package (VASP) [31] for the magnetic calculations with plane wave basis. The exchange-correlation energy were considered under Perdew-Burke-Ernzerhof (PBE) type generalized gradient approximation (GGA) [32]. For the calculation of $Cr_{1/3}NbS_2$, the GGA+U method was applied to describe the localized 3*d* orbitals of Cr atoms, where U = 4.0 eV was selected. Spin-orbit coupling was not included due to its minor effect on band structure. Experimental lattice parameters were used. The cutoff energy for the plane-wave basis was set to 480 eV for calculations with QUANTUM ESPRESSO and 400 eV for calculations with VASP. A Γ-centered Monkhorst-Pack *k*-point mesh of $9 \times 9 \times 4$ with a spacing of 0.15 Å$^{-1}$ was adopted in all self-consistent calculations.

### III.  RESULTS AND DISCUSSIONS

Figure 2 shows the comparative study of the electronic structure of $NbS_2$ and $Cr_{1/3}NbS_2$. On the FS of $NbS_2$, we observe a large hole pocket around the $\bar{\Gamma}$ and $\bar{K}$ points, respectively [Fig. 2(a)]. We do not observe the band splitting caused by the coupling between the two $NbS_2$ layers in one unit cell [33]. In $Cr_{1/3}NbS_2$, by contrast, we observe an extra small hole pocket around the $\bar{\Gamma}$ point [Fig. 2(b)]. Due to the $(\sqrt{3} \times \sqrt{3})R(30°)$ superstructure, the Brillouin zone (BZ) of $Cr_{1/3}NbS_2$ is rotated by 30° and shrinks by 2/3, compared to the BZ of $NbS_2$. Along $\bar{\Gamma}\bar{K}$ and

$\overline{\Gamma M}$, we observe only one band crossing $E_F$ in NbS$_2$, while there are two bands, marked as β and γ, crossing $E_F$ along $\overline{\Gamma K M}$ and $\overline{\Gamma M \Gamma}$ in Cr$_{1/3}$NbS$_2$ [Figs. 2(c)-2(f)].

Figures 3(a)-(h) show the band structure of Cr$_{1/3}$NbS$_2$ measured with different photon energies. Both the β and γ bands show weak $k_z$ dependence except that the spectral weight of the γ band is enhanced at high photon energies due to the matrix element effect. We observe an extra α band with its band top touching $E_F$. It shows a continuum-like spectral weight distribution, suggesting its strong $k_z$ dispersion, similar to the α band in NbS$_2$ [Fig. 2(c)]. Apparently, the electronic structure of Cr$_{1/3}$NbS$_2$ cannot be derived from a rigid shift of the electronic structure of NbS$_2$ induced by electron donation from Cr ions. Although the overall dispersion of the α and γ bands are very similar in NbS$_2$ and Cr$_{1/3}$NbS$_2$, their relative energy positions are different. The α band shifts towards $E_F$ for about 250 meV, while the γ band shifts towards higher binding energies for about 100 meV in Cr$_{1/3}$NbS$_2$. Moreover, the β band shows up only in Cr$_{1/3}$NbS$_2$, suggesting that the Cr-derived electron states contribute significantly to the electronic structure near $E_F$.

To further understand the influence of Cr atoms on the electronic structure of Cr$_{1/3}$NbS$_2$, we perform *ab-initio* calculation on the electronic structures of NbS$_2$ and Cr$_{1/3}$NbS$_2$, as illustrated in Fig. 4. For better comparison, the result of NbS$_2$ is artificially folded into a ($\sqrt{3} \times \sqrt{3}$)R(30°) super-structured BZ similar to that of Cr$_{1/3}$NbS$_2$. Figures 4(a) and 4(b) compare the electronic structures of NbS$_2$ and paramagnetic Cr$_{1/3}$NbS$_2$ (onsite Coulomb interaction U = 4 eV). Apparently, the electronic structure of Cr$_{1/3}$NbS$_2$ cannot be understood by electron doping from intercalated Cr ions. Rather, the number of energy bands crossing $E_F$ increases due to the hybridization between the Cr- and Nb-derived states, consistent with our experiment in Fig. 2. In the FM state, we observe spin-splitting of the energy bands due to the exchange interaction [Fig. 4(c)]. In Fig. 4(d), we compare the calculated electronic structure of $Cr_{1/3}NbS_2$ in the CHM and FM states. No noticeable difference is observed, suggesting the same influence of the CHM and FM ordering on the electronic structure, which is reasonable considering the large helix period of the CHM state [24,26,27]. Therefore, the DMI that is responsible for the spiral spin ordering serves as a weak perturbation to the FM state. Figure 4(e) shows the orbital-projected calculation of the electronic structure of $Cr_{1/3}NbS_2$ in the FM state. We observe substantial spectral weight of Cr $d$ orbitals near $E_F$, confirming the strong hybridization between Cr- and Nb-derived states, consistent with previous *ab-initio* calculations [24,26,27].

Figure 5 tracks the temperature evolution of the band structure of $Cr_{1/3}NbS_2$ measured with improved energy resolution using relatively low photon energy. Figures 5(a)-(e) show ARPES intensity maps at selected temperatures. In the CHM state at 15 K, we observe four bands near $E_F$, marked as α, $β_1$, $β_2$, and γ in Fig. 5(a), compared to the two bands β and γ in Fig. 2. With increasing temperature, the splitting between $β_1$ and $β_2$ gradually decreases and can hardly be resolved at 195 K. To quantify the evolution of the band structure, we collect the momentum distribution curves (MDCs) at $E_F$ taken at different temperatures [Fig. 5(f)] and fit each MDC to multiple Lorentzians. The extracted Fermi crossings of the $β_1$, $β_2$, and γ bands are overlaid on the false-color plot of the MDCs in Fig. 5(g). The $β_1$ and $β_2$ bands approach each other with increased temperature, with a sudden change of their splitting near 125 K, where the γ band also shows a sudden change of its position. From the temperature evolution of the energy bands, we conclude that the CHM ordering strongly influences the band structure of $Cr_{1/3}NbS_2$, and the splitting between the $\beta_1$ and $\beta_2$ bands are the exchange splitting caused by the CHM ordering.

In general, as described by the Stoner model, if the long-range FM order is mainly contributed by the itinerant electron around $E_F$, the magnitude of exchange splitting in metallic FM materials is proportional to the magnetic moment and becomes zero above $T_c$. In Fig. 5(h), we compare the temperature evolution of the exchange splitting and the magnetic moment. Below $T_c$, the exchange splitting follows the increasing magnetic moment with decreasing temperature, suggesting a strong interplay between the conduction carriers and the magnetic moment of the system. Above $T_c$, interestingly, the exchange splitting deviates from the magnetic moment curve and persists up to at least 195 K [Figs. 5(g) and 5(h)], in drastic contrast to the Stoner model, suggesting that $Cr_{1/3}NbS_2$ is not a band magnetic material.

The exchange splitting that is persistent above $T_c$ has been widely observed in metallic ferromagnets, including Fe, Ni, $SrRuO_3$, and $Fe_3GeTe_2$ [34-39], despite the controversial experimental results in Ni [40]. However, the temperature dependence of the exchange splitting is quite different in these materials. In $SrRuO_3$ and $Fe_3GeTe_2$, the bands stay put with increasing temperature, suggesting a minor impact of long-range magnetic ordering on the exchange splitting [37,38], consistent with a localized intra-atomic exchange interaction [34]. By contrast, the exchange splitting in $Cr_{1/3}NbS_2$ strictly follows the magnetic moment below $T_c$ but shows minor temperature dependence above $T_c$, in good consistence with the result in Ni. This behavior can be well understood by the Oguchi model that describes short-range order by exactly treating the interactions between neighbouring magnetic ion pairs while approximating the rest pairs with a mean-field [34]. The persistent exchange splitting above $T_c$ can thus be attributed to the short-range FM order at high temperatures, consistent with previous measurements of the magnetism that suggests a short-range exchange interaction in $Cr_{1/3}NbS_2$ [41,42].

## IV. CONCLUSION

In conclusion, we have systematically investigated the electronic structure of the CHM $Cr_{1/3}NbS_2$ and its temperature evolution. Both experiment result and *ab-initio* calculation show strong hybridization between Cr- and Nb-derived states near $E_F$, in contrast to a simple Cr-doping rigid shift scenario. Moreover, we observe exchange splitting of the band structure that persists far above $T_c$, suggesting a local-moment magnetism instead of band magnetism and an important role of short-range magnetic interaction in $Cr_{1/3}NbS_2$. Our results provide more insight in the understanding of the electronic and magnetic properties of the mono-axis CHM $Cr_{1/3}NbS_2$.

**ACKNOWLEDGMENTS**

This work was supported by the National Natural Science Foundation of China (Grants No. 11774190, No. 11427903, No. 11634009), the National Key R&D program of China (Grants No. 2017YFA0304600, No. 2017YFA0305400, and No. 2017YFA0402900), and EPSRC Platform Grant (Grant No. EP/M020517/1). Use of the Stanford Synchrotron Radiation Light Source, SLAC National Accelerator Laboratory, is supported by the US Department of Energy, Office of Science, Office of Basic Energy Sciences under Contract No. DE-AC02-76SF00515. This research used

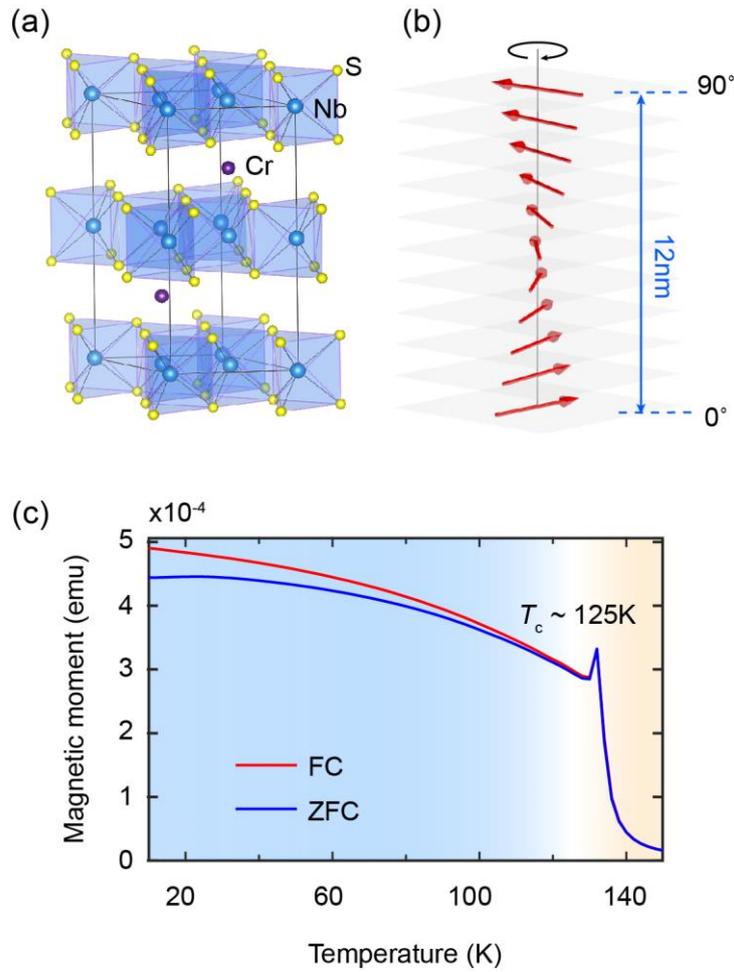

FIG. 1. (a) Crystal structure of $Cr_{1/3}NbS_2$. Cr atoms are intercalated in 2H-$NbS_2$, occupying the octahedral interstitial sites between two $NbS_2$ layers and forming a ($\sqrt{3} \times \sqrt{3}$)R(30°) superstructure. (b) Schematic illustration of the magnetic helix along *c* axis showing 1/4 period. (c) Magnetic moment as a function of temperature measured by heating the sample under a magnetic field of 100 Oe parallel to the *ab* plane. The sample was cooled down to 15 K under a magnetic field of 1 T and without magnetic field for the field cooling (FC) and zero-field cooling (ZFC) measurements, respectively.

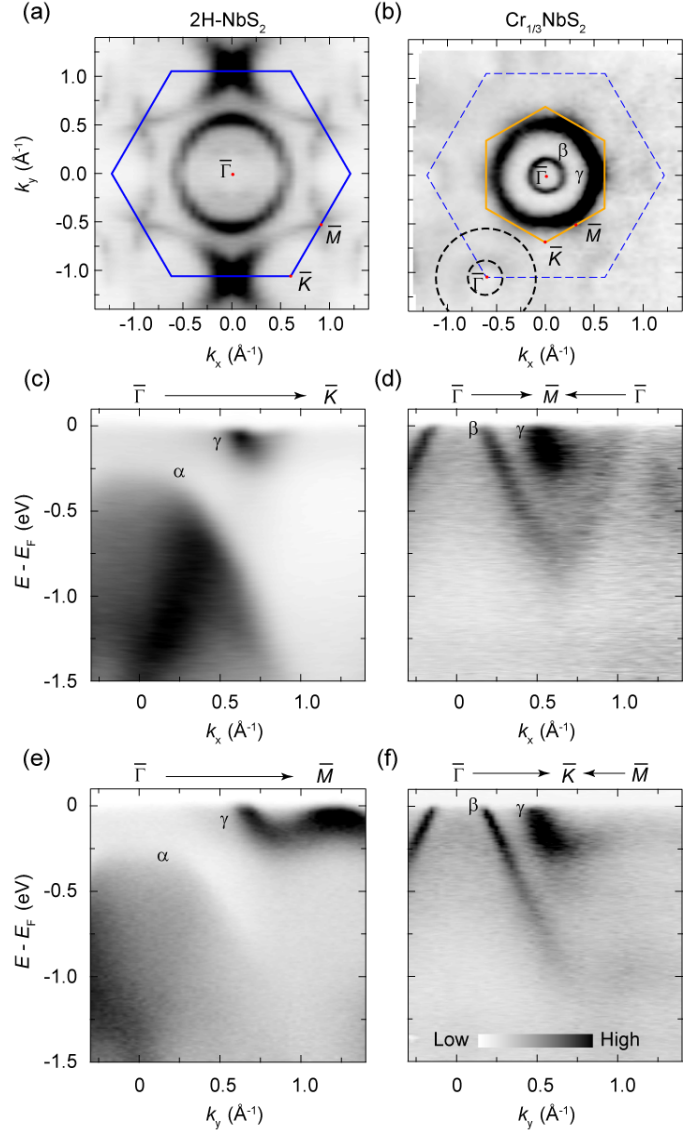

FIG. 2. (a), (b) Fermi surface of NbS$_2$ (a) and Cr$_{1/3}$NbS$_2$ (b) measured by integrating ARPES intensity over an energy window of 30 meV near the Fermi level ($E_F$). (c)-(f) Band structure of NbS$_2$ (c, e) and Cr$_{1/3}$NbS$_2$ (d, f) along high-symmetry directions as indicated. Data of NbS$_2$ (Cr$_{1/3}$NbS$_2$) were collected using 90 eV (91 eV) photons at 12 (40) K.

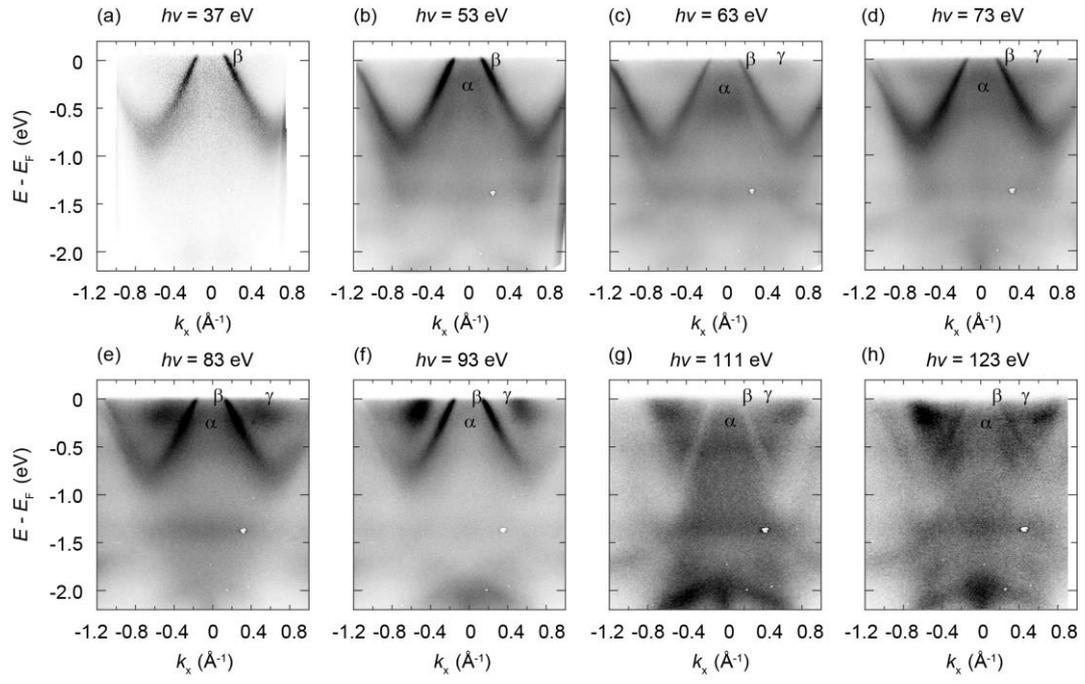

FIG. 3. (a)-(h) Band structure of $Cr_{1/3}NbS_2$ along $\overline{\Gamma}\overline{M}$ measured at selected photon energies. Data were measured at 20 K.

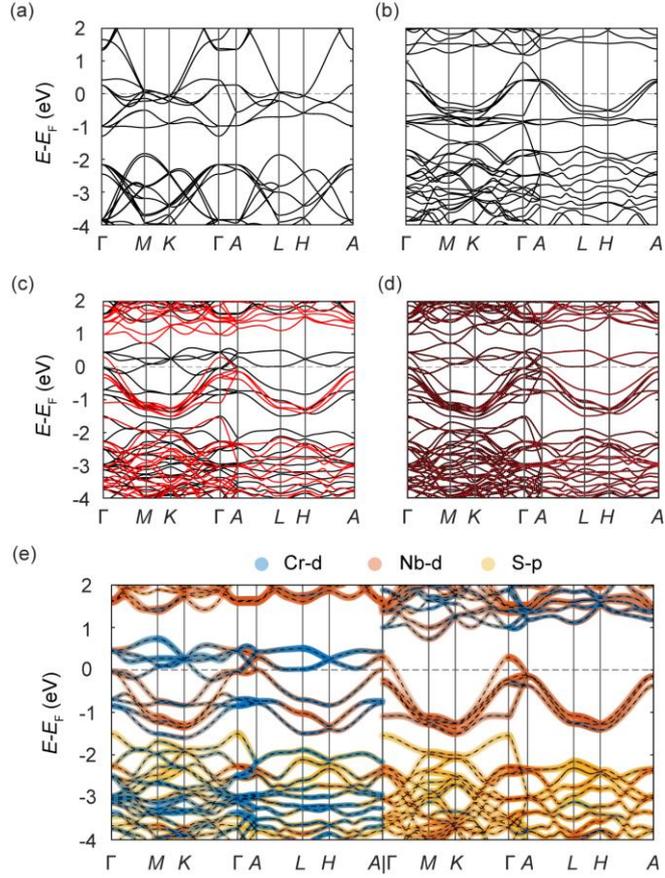

FIG. 4. *Ab-initio* calculation of the electronic structure of NbS$_2$ (a) and Cr$_{1/3}$NbS$_2$ (b-e). The calculation in (a) is folded into a ($\sqrt{3} \times \sqrt{3}$)R(30°) superstructured BZ for better comparison with the result in Cr$_{1/3}$NbS$_2$. The calculation in (b) and (c) were performed for paramagnetic and ferromagnetic (FM) states, respectively. The black (red) lines in (c) represent spin up (down) bands. (d) Comparison between the calculated electronic structure in the chiral helimagnetic (CHM) (solid black) and FM (dotted red) states showing negligible difference. (e) Orbital-projected calculation of the spin-split band structure of FM state. Left: spin up. Right: spin down.

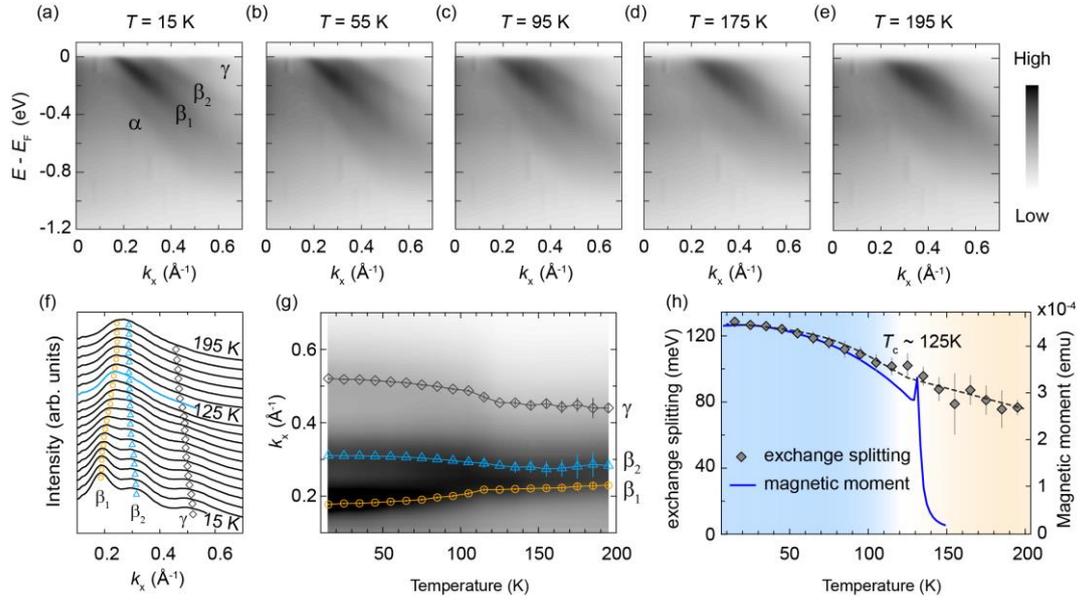

FIG. 5. (a)-(e) Band structure of $Cr_{1/3}NbS_2$ along $\bar{\Gamma}\bar{M}$ measured at selected temperatures. (f) Temperature evolution of the momentum-distribution curves (MDCs) at the Fermi energy ($E_F$). The colored markers are guides to eyes for the temperature evolution of the peak positions. (g) False-color plot of the temperature evolution of the MDCs at $E_F$. The orange circles, blue triangles, and gray diamonds indicate MDC peak positions extracted by fitting the MDCs to multiple Lorentzians. (h) The exchange splitting between the $\beta_1$ and $\beta_2$ bands as a function of temperature. The temperature evolution of the magnetic moment is also shown for comparison.